\input{epsf.tex}
\documentstyle[aps,epsf]{revtex}
\begin{document}
\title{
$^6$Li Inelastic Form Factors in a Cluster Model}
\author{S. Weber, M. Kachelrie\ss{}, M. Unkelbach and H.M. Hofmann}
\address{
Institut f\"ur Theoretische Physik,
Universit\"at Erlangen-N\"urnberg,
Staudtstr. 7
D-91058 Erlangen, Germany}
\date{\today}
\maketitle
\begin{abstract}
Longitudinal and transverse formfactors are calculated for the transition into
the low lying excited T=0 and T=1 states of $^6$Li in the framework of the
resonating group model. All formfactors are reproduced simultaneously using
three cluster-wavefunctions. Meson exchange currents yield only minor
corrections and do not lead to any specific structures at large momentum
transfers.
\end{abstract}
\narrowtext
The structure and properties of the $^6$Li nucleus are experimentally and
theoretically well studied (especially formfactors are of particular interest).
For such a light nucleus microscopic calculations, starting from a
nucleon-nucleon force, are feasible for a large variety of different models.
Special interest is devoted to three-body $\alpha$-np models often in the
framework of
Faddeev equations \cite{1,2,3,4}, but also shell model \cite{5,6}
and cluster model \cite{7,8}
calculations are affluent. In many cases only the groundstate properties of
$^6$Li are studied. We report here of an extension of ref. \cite{8}
to all low lying
excited states of $^6$Li in the framework of the
Resonating-Group-Model (RGM). We
use completly antisymmetrised RGM cluster-wavefunctions in the form $\alpha$
-np and
an effective nucleon-nucleon potential \cite{9}.

In ref. \cite{8,9} the groundstate wavefunction
of $^6$Li was calculated using Ritz
variational principle, allowing all possible combinations of S- and D-waves
on the intercluster coordinates between neutron and proton and center-of-mass
of n-p and  $\alpha$-particle. For the T=0 spin-orbit triplett ($3^+,2^+, 1^+$)
such a
variation isn't possible anymore, because the lowest energy for this partial
wave is just the $\alpha$-deuteron threshold. Therefore we used the parameters
of the
model-space of our groundstate and dia\-go\-na\-lized the Hamiltonian in the
corresponding spaces. The excitation energies are given in table \ref{tab1}.
Obviously
the agreement with experiment \cite{10} is only fair due to the fixed width
parameters. We refrained from changing the width parameters in order to
reproduce the data, because there is no controlable way to do that.

In an $\alpha$-deuteron scattering calculation
the phaseshifts for the $3^+,
2^+$ and to
lesser extend $1^+$ vary rapidly in the neighborhood of the experimental
energies
(see ref. \cite{11,12} for potentials similar to ours). For T=1
the situation is
different, since the decay into
the $\alpha$-deuteron channel is forbidden by isospin.
The $0^+$T=1 state is bound, relativ to the
$\alpha$-n-p threshold. Therefore we tried
again Ritz variational prinziple to determine the wavefunction. Besides the
obvious spin zero and pure S-wave component, we allowed P-waves on the
intercluster coordinates coupled to 1. For this model space we found a rather
stable local minimum above the 3-body breakup threshold with the parameters
$\beta=0.2838, \gamma=0.1620, \delta_1=0.2609$ and $\delta_2=0,03624$
(see ref. \cite{9} for  a explanation of these parameters).
For the $2^+$T=1 state these para\-me\-ters
were also used and we allowed all combinations of S-,P- and D-waves which
could contribute.

The energies are given below (table \ref{tab1}). Since the calculated level
ordery is not correct,
we allowed also configurations of the $^5$Li-n and $^5$He-p in the $0^+$T=1
wavefunction, gaining 0.7 MeV additional binding energy. As will be shown
later the effects of the wavefunction mo\-di\-fi\-ca\-tion on the formfactors
had
been quite small. So we won't give any details of the complicated
wavefunction.

With these wavefunctions we calculated now electromagnetic transition
formfactors. For one-body operators we used the standard expressions for the
charge density, convection current and magnetisation density \cite{8}.
For the
meson exchange currents we used the prescription of Ohta \cite{13}; details of
the calculation in the RGM framework are given in ref. \cite{14}.

Due to the finite size of the nucleons the formfactors must be modified by
the single nucleon formfactor. This is done by multiplying all matrix elements
with the nucleon formfactors attained by the well-known dipole formula
\cite{15}
(This expression is misprinted in ref. \cite{8}).

\begin{displaymath}
f(k)=\left(\frac{1}{1+\displaystyle\frac{k^2}{A_1}}\right)^{\displaystyle 2}
\mbox{with } A_1=0,71 (\rm
GeV/\hbar c)^2
\end{displaymath}

All details of the calculation are given in \cite{14}.
We mention in passing that
the groundstate wavefunction reproduces the tiny quadrupolemoment
(see \cite{9}
but also the discussion in \cite{7}) and the elastic
formfactors quite well only if
the wavefunction is properly antisymmetrised.

In Fig.\ \ref{fig1} and\ \ref{fig2} the calculated inelastic
longitudinal formfactors are displayed. The
formfactors for the T=0 states are of similar magnitude and k-dependence since
in all cases the C2 is the dominant contribution. For the $2^+$T=1 our result
is
some orders of magnitude smaller, for the $3^+$T=0 state the calculation agrees
well with the data \cite{16}. For all other
resonances there exists no data, either
due to the width of the T=0 states or the smallness of the longitudinal
formfactors for the $2^+$T=1 state. For the T=0 triplett states our results are
similar to those of ref. \cite{3}, whereas for the $2^+$T=1 we disagree in
magnitude
and form.

In Fig.\ \ref{fig3} and\ \ref{fig4} the
calculated transverse formfactors are displayed together with
the multipoles. E2- and M3-transition are the dominant ones. Whereas the
absolute magnitude agrees well with the findings of ref. \cite{3},
the results for
the various multipoles are quite different. Unfortunately there is no data to
compare with. Note that in our model meson exchange contributions are not
possible for this transition due to the isospin zero of initial and final
state.

For the T=1 states however such contributions are possible. In Fig.\ \ref{fig5}
we
compared the calculation with experimental data for the $0^+$T=1 state
\cite{16,17}.
The one-body M1 operator reproduced the data quite nicely. The MEC's are
only of minor importance, but reducing the calculated results and thus
yielding a somewhat less agreement. Therefore we allowed for additional
$^5$Li-n and $^5$He-p configurations
in the $0^+$T=1 wavefunction. The result was a
further small reduction in the formfactor similar to the MEC effects, and
even less, but still satisfactory agreement with data (see Fig.\ \ref{fig5}).
The
reduction can be easily explained by the reduced overlap of the orbital
wavefunction for the excited state with the groundstate one due to
structures missing in the groundstate. The inclusion of these structures into
the groundstate resulted in a totally wrong quadrupolemoment. Since both
wavefunctions however reproduced the data much better than ref. \cite{3} we
didn't persue this issue further (as pointed out in ref. \cite{18} the
inclusion of additional structures may in general influence substantially the
formfactors; here it resulted only in a totally wrong quadrupolemoment)
. In Fig.\ \ref{fig6} we compare transverse
formfactors of the $2^+$T=1 state with data
\cite{16,17}. Again E2 and M3 are the
dominant contributions, with MEC's playing only a minor role, but
reducing the calculation by a few percent. Both calculations agree well
with data.

In conclusion one can say that our calculation reproduced consistently the
measured elastic \cite{8} and inelastic longitudinal and transverse formfactors
for all low lying states of $^6$Li. Meson exchange contributions yielded only
minor effects and introduced no structure at larger momentum transfers.

Discussions with A. Booten are gratefully ack\-now\-ledged.

\begin{figure}
\centerline{\epsfbox{LongFFA.eps}}
\caption{Longitudinal formfactors for the 1$^{+\ast}$T=0 and 2$^+$T=0 state}
\label{fig1}
\end{figure}

\begin{figure}
\centerline{\epsfbox{LongFFB.eps}}
\caption{Longitudinal formfactors for the 2$^+$T=1 and 3$^+$T=0 state\newline
data: Bergstrom, J.C. ...; Nucl. Phys. {\bf A327} (1979) 439}
\label{fig2}
\end{figure}

\begin{figure}
\centerline{\epsfbox{T0A.eps}}
\caption{F$_{\rm T}^2$ for the 1$^{+\ast}$T=0 and 3$^+$T=0 state}
\label{fig3}
\end{figure}

\begin{figure}
\centerline{\epsfbox{T0B.eps}}
\caption{F$_{\rm T}^2$ for the 2$^+$T=0 state}
\label{fig4}
\end{figure}

\begin{figure}
\centerline{\epsfbox{0T1.eps}}
\caption{F$_{\rm T}^2$ for the 0$^+$T=1 state\newline
full line: impulse-approximation\newline
dotted line: including MEC's\newline
data:\newline
Bergstrom, J.C. ...; Nucl. Phys. {\bf A327} (1979) 439\newline
Bergstrom, J.C. ...; Nucl. Phys. {\bf A251} (1975) 401}
\label{fig5}
\end{figure}

\begin{figure}
\centerline{\epsfbox{2T1.eps}}
\caption{F$_{\rm T}^2$ for the 2$^+$T=1 state\newline
data: Bergstrom, J.C. ...; Nucl. Phys. {\bf A327} (1979) 439}
\label{fig6}
\end{figure}

\clearpage
\begin{table}
\begin{tabular}{lll}
        \bf state:&\bf	calculated:&\bf	experimental:\\
	&&\\
	$1^+$T=0\tablenotemark[1]&	0&	0\\
	$3^+$T=0&	3.33	&2.18\\
	$0^+$T=1&	6.16, 4.46\tablenotemark[2]&3.56\\
	$2^+$T=0&	4.86&	4.31\\
	$2^+$T=1&	8.52&	5.37\\
	$1^{+\ast}$T=0&	5.37&	5.65\\
\end{tabular}
\caption{$^6$Li Energies (in MeV)\label{tab1}}
\tablenotetext[1]{groundstate}
\tablenotetext[2]{complex wavefunction}
\end{table}

\end{document}